# Real-Time Foreign Object Recognition Based on Improved Wavelet Scattering Deep Network and Edge Computing

Zhichao He, Xiangyu Shen, Yong Zhang, and Nan Xie

*Abstract*—The increasing penetration rate of new energy in the power system has put forward higher requirements for the operation and maintenance of substations and transmission lines. Using the Unmanned Aerial Vehicles (UAV) to identify foreign object in real time can quickly and effectively eliminate potential safety hazards. However, due to the limited computation power, the captured image cannot be real-time processed on edge devices in UAV locally. To overcome this problem, a lightweight model based on an improved wavelet scatter deep network is proposed. This model contains improved wavelet scattering network for extracting the scatter coefficients and modulus coefficients of image single channel, replacing the role of convolutional layer and pooling layer in convolutional neural network. The following 3 fully connected layers, also constituted a simplified Multilayer Perceptron (MLP), are used to classify the extracted features. Experiments prove that the model constructed with biorthogonal wavelets basis is able to recognize and classify the foreign object in edge devices such as Raspberry Pi and Jetson Nano, with accuracy higher than 90% and inference time less than 7ms for 720P (1280×720) images. Further experiments demonstrate that the recognition accuracy of our model is 1.1% higher than YOLOv5s and 0.3% higher than YOLOv8s.

## 1 Introduction

THE increasing penetration rate of new energy in the power system has put forward higher requirements for the operation and maintenance of substations and transmission lines. The safe and stable operation of transmission lines is a necessary guarantee for the development of social production and people's livelihood (Panahi et al., 2021). Various foreign objects on transmission lines may cause problems such as short circuits, conductor breaks and line oscillations, which bring potential threats to the electric power system; therefore, regular inspection has become an important task to guarantee the continuous supply of electric power and the safe operation of transmission lines (Liu et al., 2022). With the development of automation and robotics technology, human inspection gradually being replaced by Unmanned Aerial Vehicles (UAV) inspection or wheeled/legged robot inspection (Nguyen et al., 2019), adopting the "UAV inspection as the main, manual inspection as a supplement" approach has been developed for transmission line inspection in the main operation and maintenance module (Ma et al., 2021); on the other hand, the technology based on wheeled or quadruped-driven inspection robots is also gradually maturing. Recognizing the huge number of UAV inspection images with human naked eyes is not only time-consuming, but also frequently leads to misjudgments (Odo et al., 2021). While centralized intelligent recognition of inspection images through computer algorithms has the advantages of high accuracy and fast detection speed (Souza et al., 2023).

At this stage, foreign object recognition methods are mainly divided into two main categories. One category is based on traditional foreign object recognition methods. Wang et al. (2015) proposed a method for detecting broken strands and foreign object defects in transmission lines based real-time structure sensing, which can identify broken wires and foreign object defects by calculating the width variation of segmented conductors and the gray scale similarity. Jiao and Wang (2016) proposed a frame difference labeling method, which utilizes key frame extraction and foreign object feature point tracking to achieve the purpose of transmission line foreign object detection. Haroun et al. (2021) proposed the use of satellite images to improve the feature information of Support Vector Machine (SVM), thus improving the efficiency of foreign object detection. Ye et al. (2020) proposed the use of particle swarm algorithm to optimize SVM for foreign object detection. Traditional methods suffer from low detection accuracy, single-target detection, poor generalization ability and poor scalability (Liu and Wu, 2023).

The other category is based on deep learning methods. The Convolutional Neural Network (CNN) structure has become a widely used method in the field of transmission line defect detection (Tian et al., 2022). Wang et al. (2017) compared and analyzed three methods: Deformable Part Model (DPM), Faster R-CNN, and Single Shot MultiBox Detector (SSD), using real datasets of transmission line foreign objects to verify the feasibility of deep learning-based recognition methods in real-time detection of foreign objects. Qiu et al. (2023) proposed a lightweight YOLOv4 model embedded with a dual attention mechanism, utilizing MobileNetV2 for feature extraction, as well as depth-separable convolution instead of standard convolution in the SPP and PANET modules, and embedded with a convolutional block-attention module to improve the detection accuracy. Sun and Yi (2024) proposed to replace the backbone of YOLOv7-tiny with a ReXNet network, which introduces diverse branching blocks and dramatically improves the detection efficiency. Singh et al. (2023) conducted a comparative performance analysis of multiple models from YOLOv5 to YOLOv8, which verifies the advantages of the YOLO series in the detection of foreign objects on transmission lines.

Although CNN-based recognition algorithms are higher in accuracy than traditional image algorithms, they consume large amounts of computing power, need expensive computer devices such as Graphic Computer Unit (GPU), and usually need to upload inspection images to a cloud server for detection, which does not allow for in situ real-time monitoring (Hu et al., 2024). In order to meet the real-time requirements, the use of miniaturized convolutional neural network structures and edge computing devices to detect the images in the field is gradually gaining attention (Zhou et al., 2020; Hu et al., 2021). Hu et al. (2024) proposed a diagnosis method for defects of transmission lines based on LEE-YOLOv7, which introduces the LCnet network and invokes the TensorRT module, resulting in a significant increase in both detection accuracy and speed. Liu et al. (2020) proposed to build an edge computing system with Nvidia TX2, which greatly reduces the amount of data transmission and improves the efficiency of inspection compared with the solution of transmitting data back to the server. Yu et al. (2022) proposed to use Otsu algorithm to extract the target region, then extract its depth features by DenseNet201, and finally utilize ECOC-SVM algorithm for training and testing to improve the accuracy of foreign object detection. Liu et al (2024) improved YOLOv5 and then pruned it by ResRep to achieve a significant increase in inference speed.

The above methods have great accuracy due to the outstanding feature extraction ability of CNN, but the requirement for computing power is still very high. Even after optimization and pruning of the model, it is difficult to balance the inference accuracy and real-time performance with edge computing. On the other hand, the cloud computing solution has some problems brought by wireless image transmission, such as communication interruption, resolution limited by bandwidth, and time delay (>70ms) caused by long-distance transmission. Up to now, the research on high-precision (>90%), real-time (≤33ms) foreign object recognition for UAV generated High Definition (HD) whose resolution is 1920×1080 with 30 Frame Per Second (FPS) videos is still limited.

In this paper, an improved wavelet scattering based deep network is proposed to realize the real-time recognition of foreign objects in transmission lines, especially for power systems with high penetration of new energy. To meet the real-time requirement of detection, the structure of wavelet scattering network is improved for supporting arbitrary layers. The biorthogonal wavelets basis including bior1.1, bior2.2, and bior1.3 wavelets are utilized to construct a three-layer scattering network, replacing the convolutional layer and pooling layer in the traditional CNN model. The new networks can effectively extract the key features of the foreign objects, simplify the network structure and reduce the model's computational complexity. The modulus coefficients and scattering coefficients in scattering network are selected for better characterizing of the foreign object, as the input of a 3-layer small deep network for recognition. The algorithm proposed in this paper shows strong generalization ability, and its real-time detection of transmission line equipment defects can also be realized by choosing appropriate wavelet base and scattering value features. Experimental results show that the method can effectively reduce the amount of model computation, and local real-time detection can be realized on edge devices with accuracy larger than 94% and a process speed of 149.3 FPS.

## 2 Analysis of convolutional neural network inference speed

Fig. 1 shows the diagram of two UAV intelligent inspection models. One is based on the Centralized Cloud computation, where the inspection video or images are transmitted through wireless communication. Due to the bandwidth limitation, the transmitted real-time video usually has limited resolution (<1080P) and large compression ratio and may lose some key figure details, thus it is not suitable as evidence for foreign object judgment. Usually, high resolution 4K of 8K photo image with clear detail is transmitted through UAV instead of video. For distance larger than 2km, the delay time of HD image transmission is usually about 100 microseconds without communication interference. The advantage of the cloud computation is the massive computing resources. The popular Yolo model is deployed as the recognition unit and achieve high precision judgment. The detection algorithm can be iteratively updated at any time. The disadvantage is listed as follows: 1) it is not a real-time detection method; the response is not fast enough to compete with the image acquire speed. 2) The image transfer and judgement processes always use the off-line module, the image is acquired and collected, and then transmitted and processed together for higher efficient, but they are unable to respond smartly based on the on-site situation. 3) The wireless transfer process is sensitive to interference. For extreme harsh environment, e. g. the thunderstorm weather, the wireless transmission may suffer from image information loss or larger transmitted delay, and even the transfer failure of entire image. 4) High power consumption. The wireless image transmission costs large amount of power and reduces the inspection time, while the implement of recognition in GPU also consumes huge power. Thus, the cloud computation model is not energy conservative and environmentally friendly.

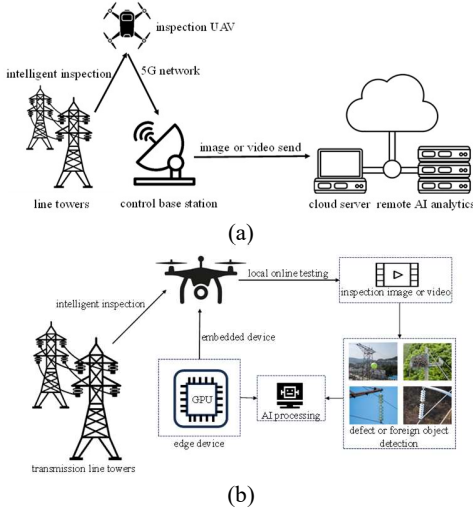

**Fig. 1** UAV intelligent inspection system. (a) the cloud computing-based intelligent inspection system for UAV. (b) intelligent inspection system for UAV based on edge computing.

Another model is using the edge computation with in-situ processing within UAV itself, as illustrated in Fig. 1b. The edge computation requires low energy, and the transfer process of the total image could be omitted. The in-situ edge computation is possible to achieve real-time on-line recognition for foreign objects, supplying the smart timely response for sudden special circumstances. It is helpful to realize the flexible inspection process according to on-site situation instead of following a fixed inspection route, thus improve the inspection efficiency dramatically.

However, the model based on edge computation requires that the recognition algorithm has very high efficiency, since the edge devices' computing power is very limited compared to central AI computation server. For real-time detection of the UAV's inspection video, the algorithm's efficient should be outstanding, guaranteeing the accuracy and processing speed at the same time. Unfortunately, the popular CNN model relatively consumes large computer power, and is difficult to realize edge computation for HD videos.

Currently, there are multiple challenges in the field of transmission line foreign object recognition. First, it is difficult to identify foreign objects in a uniform manner due to the diversity of the foreign objects themselves such as their different shapes, sizes and materials; Second, the complexity of the recognition environment such as background interference, illumination changes, and dynamic environments substantially increases the difficulty of recognition; Finally, for certain application scenarios, such as real-time monitoring and detection in power systems, the real-time nature of the recognition is strictly required. Although existing techniques can localize and identify foreign objects with some degree of accuracy, there are still significant performance limitations when dealing with real-time detection tasks on edge devices. The reason for this is mainly the huge computational complexity that constrains the operational efficiency, which is analyzed as follows:

The computational complexity is the most commonly used metric to evaluate the size of a CNN model, also known as the number of Floating-Point Operations (FLOPs). Assuming that the input image size is $N \times L$, the output size of the convolutional and pooling layers is represented as:

$$M_1 = \frac{N - D(K-1) - 1 + 2P}{S} + 1 \quad (1)$$

$$M_2 = \frac{L - D(K-1) - 1 + 2P}{S} + 1 \quad (2)$$

where $N$ and $L$ are the size of the input image's side length, $K$ is the convolutional kernel size, $P$ is the padding, and $S$ is the stride and dilation factor $D$ (generally defaulted to 1). Combining (1) can get the formula for the computational complexity of the convolutional layer:

$$FLOPs = M_1 \times M_2 \times (K^2 \times C_{in} + 1) \times C_{out} \quad (3)$$

where $C_{in}$ is the input channel and $C_{out}$ is the output channel. Inside the parentheses is the sum of the multiplication and addition operands, the above equation considers the bias, if not, +1 will be neutralized.

The Fully Connected (FC) layer is considered as a special convolutional layer, where the size of the convolutional kernel is equal to the size of the input image, and the formula for the computational complexity of the FC layer can be expressed as:

$$FLOPs = (I+1) \times O \quad (4)$$

where $I$ is the number of input neurons and $O$ is the number of output neurons. The above equation is again considering bias and there is no +1 when it is not considered.

When the pooling layer is maximum pooling, there is not a complete floating-point multiply and add computation in this layer, and together with the small amount of computation in the pooling layer, it can be considered as 0. When it is average pooling, its computational complexity is given by:

$$FLOPs = C_{in} \times W_{in} \times H_{in} \times K^2 \quad (5)$$

where $W_{in}$ and $H_{in}$ are the width and height of the input image and $K$ is the pooling window size.

The computational complexity of the ReLU activation function is equal to the size of the input samples, and in the case of deep learning models, the activation function layer is negligible in comparison to the convolutional and fully connected layers.

Taking a simpler convolutional neural network as an example, its structure is $Pool_{ave}(ReLU(Conv2))$, $ReLU(FC)$, $ReLU(FC)$, which can be understood as a three-layer structure, where $Pool_{ave}$ denotes the average pooling, and $Conv2$ denotes the two-dimensional convolution. Combining (1), (2) and (3), we can get the computation of the first layer:

$$FLOPs = (\frac{N - D(K-1) - 1 + 2P}{S} + 1) \times$$
$$(\frac{L - D(K-1) - 1 + 2P}{S} + 1) \times \quad (6)$$
$$(K^2 \times C_0 + K_1^2 + 2) \times C_1$$

where $N$ and $L$ are the size of the input image's side length, $K$ is the convolution kernel size, $P$ is the padding, $S$ is the stride and dilation factor $D$, $K_1$ is the pooling kernel size, $C_0$ is the input channel and $C_1$ is the output channel.

Similarly, the computation of the last two layers can be obtained:

$$FLOPs = \left\{\frac{[M_1 - D_1(K_1-1)-1+2P_1]}{S_1}+1\right\} \times$$
$$\left\{\frac{[M_2 - D_1(K_1-1)-1+2P_1]}{S_1}+1\right\} \times \quad (7)$$
$$O_0 + 2O_0 + (O_0+2) \times O_1$$

where $M_1$ and $M_2$ are the convolutional output edge size, $P_1$ is the padding, $S_1$ is the stride and dilation factor $D_1$, $O_0$ and $O_1$ are the number of output neurons in the first FC layer and the second FC layer respectively.

**Table 1 Calculations for the same structure with 3 resolution inputs**

| Size | GFLOPs |
|---|---|
| 960×540 | 0.46 |
| 1280×720 | 0.82 |
| 1920×1080 | 1.85 |

If the input sizes of color images is 960×540 (540P),1280×720 (720P) and 1920×1080 (1080P), assuming that the convolution kernel is 7×7, pooling kernel is 5×5, the padding is 0, the stride is 1, the number of channels is unchanged, the second layer output neuron is 128, the third layer output neuron is 64, the computational complexity of the three resolutions can be calculated by (2), as shown in Table 1.

As shown in Table 1, although the network only contains one convolutional layer, the computational complexity for images of 540P resolution still reaches 0.46G FLOPs. However, the current popular large models such as YOLOv8 contain nearly ten convolutional layers, and the computational complexity of the model exceeds 10GFLOPs. With the current computing power of ARM or RISC-V devices, their inference time will reach seconds, which is more than the video stream processing requirement of 30 FPS for drone cameras, making it difficult to meet the requirement of real-time image inference speed. Therefore, it is urgent to develop more efficient algorithms, optimize the network structure of the model, and achieve fast and accurate identification of foreign objects on transmission lines.

## 3 Improved wavelet scattering network principle

In this paper, the Improved Wavelet Scattering Network (IWSN) is used to replace the convolutional layer and pooling layer to realize the feature extraction of the foreign object. The modulus coefficients and scattering coefficients by IWSN are used as the inputs, which can greatly reduce the input information. IWSN can effectively solve the contradiction between accuracy and inference time; and realize high-precision and fast identification of foreign object of the contact network on the edge intelligent devices.

### 3.1 Principles of Wavelet Scattering Networks

Wavelet scattering network is based on the scattering operator of wavelet transform, which can extract the invariant properties of the image information in the scattering transform, and its structure is similar to the CNN, the difference is that its filter is a pre-determined wavelet filter, and it does not need to learn to obtain the parameters of the network through the training samples. Wavelet scattering network can extract high-frequency information and low-frequency information features hierarchically and perform nonlinear modulo operation, and the obtained signal feature expression has excellent properties such as translation invariance and deformation stability, which precisely meets the basic requirements of feature extractors in machine learning (Mallat, 2012; Andén and Mallat, 2014; Gowthaman and Das, 2025).

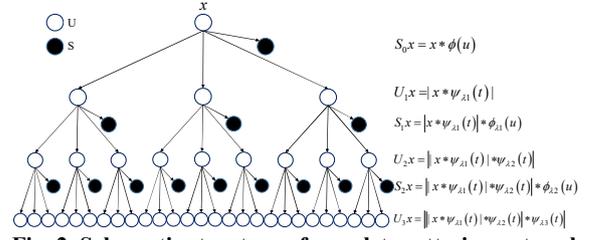

**Fig. 2 Schematic structure of wavelet scattering network**

Fig. 2 shows the process of constructing a wavelet scattering network. In this process, the input signal, denoted as $x$, is passed through the first layer of the network structure to produce the scattering coefficient $S_0$ and the modulus coefficient $U_1$, where $U_1$ becomes the input to the next layer. Then, the second layer of the network structure generates the scattering coefficient $S_1$ and the modulus coefficient $U_2$, where $U_2$ becomes the input to the third layer of the network structure, and so on. In this way, the signal is processed layer by layer, with new coefficients generated at each step (Wiatowski and Bölcskei, 2018; Wang et al., 2023).

The specific calculation formula is as follows:
The 0$th$ order scattering coefficient $S_0$ is expressed as:
$$S_0 x = x * \phi(u) \quad (8)$$

where $S_0$ is the 0$th$ order scattering operation operator, $x$ is the input 2D image information, $*$ is the convolution symbol, and $\phi(u)$ is the scale function of any of the preset wavelet bases.

To obtain the input signal for the first layer, the original signal $x$ is convolved with the wavelet function $\psi_{\lambda_1}(t)$, where $\lambda_1$ is the used wavelet label and $t$ is the variable of wavelet function, and then modulo operations $U_1$ are performed to obtain:

$$U_1 x(u, \lambda_1) = |x * \psi_{\lambda_1}(t)|. \quad (9)$$

Convolving this signal with the scale function $\phi_{\lambda_1}(u)$ yields the 1$st$ scattering coefficient:

$$S_1 x(u, \lambda_1) = |x * \psi_{\lambda_1}(t)| * \phi_{\lambda_1}(u) \quad (10)$$

where $S_1$ is the 1$st$ order scattering operation operator.

Iterating layer by layer, the modulus coefficient and the scattering coefficient of the n$th$ layer with wavelets: $\lambda_1, \lambda_2,\ldots, \lambda_n$ are:

$$U_n x(u,\lambda_1,\cdots,\lambda_n) = \left|\cdots\left\| x*\psi_{\lambda 1}(t)\right|*\psi_{\lambda 2}(t)\right|\cdots*\psi_{\lambda n}(t)\right| \\ = \left|U_{n-1}x(u,\lambda_1,\cdots,\lambda_{n-1})\psi_{\lambda n}(t)\right| \quad (11)$$

$$S_n x(u,\lambda_1,\cdots,\lambda_n) \\ = \left|\cdots\left\| x*\psi_{\lambda 1}(t)\right|*\psi_{\lambda 2}(t)\right|\cdots*\psi_{\lambda n}(t)\right|*\phi_{\lambda n}(u) \quad (12)\\ = U_n x(u,\lambda_1,\cdots,\lambda_n)*\phi_{\lambda n}(u).$$

### 3.2 Improved Wavelet Scattering Network

The number of network layers of the classical wavelet scattering network is determined by the number of wavelet bases in the preset filter, which limits the flexibility of the network structure. In this paper, the classical wavelet scattering network structure is improved so that the number of network layers can be unrestricted by the number of wavelet bases, and a small number of wavelet bases can be used for multilayer decomposition of high-resolution images.

The modulus coefficients and scattering coefficients of IWSN are defined as follows:

The 0$th$ order scattering coefficient $S_0$ is also expressed as:

$$S_0 x = x*\phi(u). \quad (13)$$

The 1$st$ order modulus coefficients and scattering coefficients are defined as:

$$U_1 x(u,\lambda_1) = |x*\psi_{\lambda 1}(t)| \quad (14)$$

$$S_1 x(u,\lambda_1) = U_1 x(u,\lambda_1)*\phi_{\lambda 1}(u). \quad (15)$$

The 2$nd$ layer of $U_2$ and $S_2$ can be expressed as:

$$U_2 x(u,\lambda_1,\lambda_2) = \left\| x*\phi_{\lambda 1}(u)\right|*\psi_{\lambda 2}(t)\right| \quad (16)$$

$$S_2 x(u,\lambda_1,\lambda_2) = U_2 x(u,\lambda_1,\lambda_2)*\phi_{\lambda 1}(u). \quad (17)$$

And so on, the m$th$ layer's modulus coefficient and scattering coefficient are:

$$U_m x(u,\lambda_1,\cdots,\lambda_m) = \left|\cdots\left\| x*\phi_{\lambda 1}(u)\right|\cdots*\phi_{\lambda m-1}(u)\right|*\psi_{\lambda m}(t)\right| \quad (18)$$

$$S_n x(u,\lambda_1,\cdots,\lambda_n) = U_n x(u,\lambda_1,\cdots,\lambda_n)*\phi_{\lambda 1}(u). \quad (19)$$

The IWSN redefines the modulus coefficients, and unlike the classical modulus coefficients, the positions of the scale function and the wavelet function are interchanged in the expression.

For the 2D image, the low frequency coefficients and the high frequency diagonal coefficients are selected as the convolution kernel of the IWSN, and the two expressions are:

$$\phi(x_1,x_2) = \sum_{n1}\sum_{n2} h(n_1)h(n_2)\phi(2x_1-n_1, 2x_2-n_2) \quad (20)$$

$$\psi(x_1,x_2) = \sum_{n1}\sum_{n2} g(n_1)g(n_2)\phi(2x_1-n_1, 2x_2-n_2) \quad (21)$$

where $\phi(x_1,x_2)$ is the scale function, $\psi(x_1,x_2)$ is the wavelet function, $h$ is the low-pass filter, $g$ is the high-pass filter, and $n_1$ and $n_2$ are the length and width of the input image.

## 4 Design and Optimization of Improved Wavelet Scattering Depth Networks

### 4.1 Basic Structure of Combinatorial Models

The key features extracted by the wavelet scattering network can be utilized for identification and classification using a small deep network. Thus, a combined model based on IWSN and multilayer deep network is employed to realize fast and high accuracy recognition of foreign objects. The basic structure of the improved wavelet scattering depth network is shown in Fig. 3b. Firstly, a color channel of the foreign object RGB image is extracted, and then it is utilized as the input of the IWSN, the structure of the IWSN is shown in Fig. 3a, and its outputs are the modulus coefficient $U$ and the scattering coefficient $S$. The specific $U$ and $S$ are combined and fed into a multilayer deep network structure. In order to reduce the inference time, the deep network contains only fully connected and activation layers, and the total number of layers is limited to ten to reduce the computation consumption.

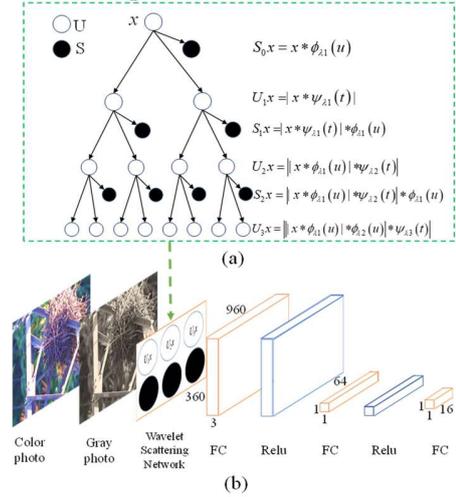

**Fig. 3 The basic structural diagram of the improved wavelet scattering depth network**

In order to reduce the amount of information in the input image, only one of the color channels of RGB image is used in this paper. In order to verify which color channel of R channel, G channel and B channel has the best effect, this paper conducts a comparison experiment. Firstly, the 3 color channels of the picture are extracted; then the corresponding modulus coefficient $U$ and scattering coefficient $S$ are obtained by inputting the wavelet scattering network respectively, in which the wavelet bases used in each layer of the scattering network are bior1.1; finally, the three-layer neural network is used to identify and classify each channel. The specific experimental data are shown in Table 2.

**Table 2 Comparison of recognition accuracy results of different channels for extracting each foreign object picture**

| Type of object | R channel accuracy (%) | G channel accuracy (%) | B channel accuracy (%) |
|---|---|---|---|
| bird's nest | 94.8 | 94.8 | 94.8 |
| kite | 93.8 | 93.8 | 93.8 |
| textile | 92 | 92 | 92 |
| plastic | 91 | 91 | 91 |

As can be seen from Table 2, the recognition accuracies of the three channels are identical under the same conditions. For the convenience of presentation, we decide to use the B color channel in the foreign object image as the input.

**4.2 Design of Improved Wavelet Scattering Networks**

The biorthogonal wavelet basis is a set of paired wavelet basis functions which are used for signal decomposition and reconstruction respectively. Unlike orthogonal wavelets, there is orthogonality between the wavelet functions of biorthogonal wavelets at different scales of scaling, while there is no orthogonality between those at the same scale. Therefore, the wavelets used for decomposition and reconstruction are not the same function, and this design solves the problem of incompatibility between symmetry and exact signal reconstruction, which makes biorthogonal wavelets perform well. The family of biorthogonal functions is usually expressed in the form biorN$r$.N$d$, where the 'N$r$' denotes the reconstruction order and 'N$d$' denotes the decomposition order. In addition, the biorthogonal wavelets have several advantages such as linear phase properties benefit to maintain the phase information of the signal, regularity and tight support.

1) Selection of Wavelet Scattering Network Feature

There are wide varieties of foreign objects on transmission lines and the background of the picture is also diverse. When UAV is taking aerial photographs, the relative sizes of the foreign objects in the captured pictures may also change due to the different distances of the camera from the target object (Luo et al., 2024) To cope with the above challenges, we utilize the modulus coefficients and scattering coefficients to extract the feature of different types of foreign objects in various sizes. These two coefficients are functionally comparable to the convolutional and pooling layers in CNN.

Fig. 4 illustrates the results of the IWSN for four types of foreign objects, and the wavelet bases used are bior1.1. As can be seen from Fig. 4, the characteristics of foreign objects can be represented by the modulus coefficients and scattering coefficients calculated by the IWSN.

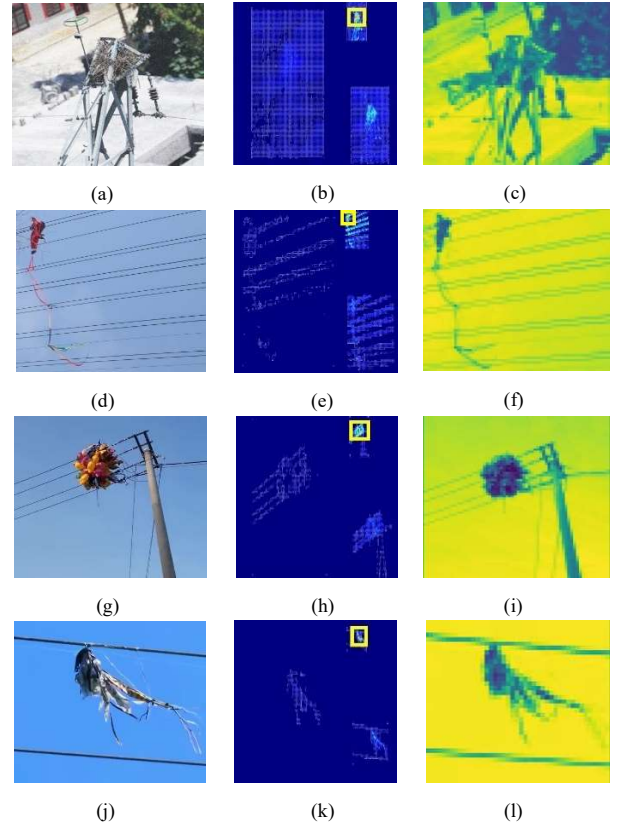

Fig. 4 Results of the IWSN for different foreign objects. (a), (d), (g) and (j) are the original image of the foreign object. (b), (e), (h) and (k) are the sets of modulus coefficients. (c), (f), (i) and (l) are the higher-order scattering coefficients.

2) Selection of Wavelet Bases and Design of Network Structure

In this paper, the biorthogonal series wavelets are adopted as the wavelet basis in the improved scattering wavelet deep network. To select the suitable wavelet basis from this series, a performance test was conducted, and the results are shown in Fig. 5.

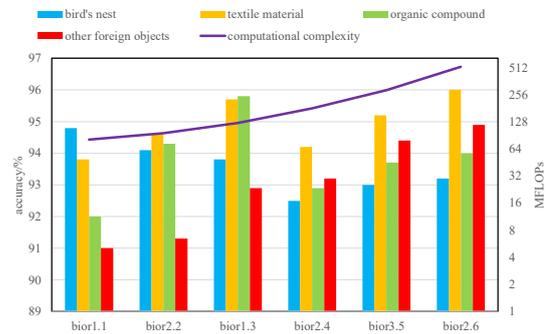

**Fig. 5 Comparison of model performance based on six different wavelet bases**

As can be seen from Fig. 5, our proposed model consumes very little computational resources, and the total computational complexity is minimized when using bior1.1 wavelet with only 81.1 MFLOPs; The total computational complexity is also less than 150 MFLOPs when bior2.2 or bior1.3 wavelets are used, whereas the total computational complexity is the highest when bior2.6 wavelet is applied, reaching 521.6 MFLOPs, six times that of using bior1.1. In terms of recognition accuracy, the bior1.1 wavelet does the best for bird's nests, reaching 94.8%; the bior2.6 wavelet has the highest recognition accuracy for textiles and other foreign objects, reaching 96% and 94.9%, respectively. But its inference speed is the slowest; the bior1.3 wavelet has the highest recognition accuracy for polymer waste such as plastic bags and waste balloons, reaching 95.8%.

To summarize, the recognition speeds of the bior1.1, bior2.2 and bior1.3 wavelets are faster and the bior1.3 wavelet has the highest total accuracy in recognizing all the foreign objects. Therefore, we choose the bior1.1, bior2.2 and bior1.3 wavelets as the wavelet bases.

Based on the above analysis, we choose the modulus coefficients $U(\lambda_1, \lambda_2, \lambda_3)$, where $\lambda_1$ is bior1.1, $\lambda_2$ is bior2.6, and $\lambda_3$ is bior1.3. $\lambda_1$ is the most efficient bior1.1 wavelet due to the largest amount of input information; $\lambda_3$ has the highest accuracy but the slowest computation speed, so the bior1.3 wavelet serves as a wavelet basis of the third order, and the size of the input information at this time is only 1/64 of the original information. Higher order scattering coefficients $S(\lambda)$ are all chosen to be bior1.1 wavelets for inference efficiency.

### 4.3 Design of Deep Classification Networks

Deep networks contain only the fully connected layer and the activation function, both of which are defined below:

The FC layer multiplies the input tensor with the weight matrix and then adds a bias to produce the output as a linear transformation with the expression:

$$\mathbf{Y}_{i \times k} = \mathbf{X}_{i \times j} \mathbf{W}_{j \times k} + \mathbf{b} \quad (22)$$

where the input is $\mathbf{X}$, the output is $\mathbf{Y}$, $\mathbf{W}$ is the parameter to be learned by the model, $\mathbf{b}$ is the $k$-dimensional vector bias, $i$ is the number of rows of the input vector, $j$ is the number of input neurons, and $k$ is the number of output neurons.

The activation function adopts ReLU function to introduce a nonlinear transformation into the neural network. Its input values are aligned with the output of the previous FC layer. ReLU function is formulated as follows:

$$f(x) = \max(0, x). \quad (23)$$

The total computational complexity of the network structure can be expressed as:

$$FLOPs = \sum_{L=k_1}^{d_1} (I_L + 1) \times O_L + \sum_{N=k_2}^{d_2} I_N \quad (24)$$

where $L$ and $N$ are the indices of fully connected and activated layers respectively, $d_1$ and $d_2$ are the depths, i.e., the number of fully connected layers and activated function, $I_L$ and $I_N$ are the inputs of the $L$th and $N$th layers, and $O_L$ is the output of the $L$th layer.

The complexity of the neural network and the recognition accuracy are directly related. According to the results of many experiments, the final actual selected network structure is shown in Fig. 3b. This deep network structure contains three FC layers and two ReLU activation functions. The three FC layers have neural unit counts transitioning from 960*360*3 to 64, then from 64 to 16, and finally from 16 to 16, the detail structure and parameters of this Multilayer Perceptron (MLP) are listed in Table 3. The first FC layer's input is 3×360×960, corresponding to the three orders modulus coefficients. Combined with (23) and (24), the expression for the computational complexity of the network structure can be obtained as:

$$FLOPs = I \times O_1 + O_1 + (O_1 + 1) \times O_2 \\ + O_2 + (O_2 + 1) \times O_3 \quad (25)$$

where $I$ is the number of input neurons and $O_1$, $O_2$, and $O_3$ are the number of output neurons in the three FC layers, respectively.

Substituting these parameters into (23-25), the computational complexity of the neural network can be calculated to be 66.3 MFLOPs, nearly equal to FLOPs of the first layer since FLOPs of other layers could be neglected, as shown in Table 3. With the computational power of the existing edge devices, the theoretical inference time of our algorithm can reach milliseconds, which is faster than the commonly used video frame rate and meets the speed requirement of real-time image inference.

Table 3 Structures and definition of the 3-layer MLP

| Layer | input | Output | FLOPs |
| --- | --- | --- | --- |
| Linear | 960*360*3 | 64 | 66.3M |
| ReLu | - | - | 64 |
| Linear | 64 | 16 | 1040 |
| ReLu | - | - | 16 |
| Linear | 16 | 16 | 68 |

## 5 Experimentation and Analysis

In order to test our algorithm's real-time performance and recognition accuracy, the collected foreign objects images are applied as the training and testing dataset, and the model has been running on the computer server and different edge devices with low power ARM CPU.

### 5.1 Experimental condition and composition of Data Sets

Experiments are conducted on computer server with 64-bit Windows 10 and edge devices with Linux Ubuntu 20.04. The hardware configuration includes a 10-core Intel(R) Core (TM) i5-10500 CPU, a 12GB NVIDIA GeForce RTX 3060 GPU, and edge computing devices such as Jetson Nano, Raspberry Pi 4B, and so on. The programming language used is Python 3.10, and the deep learning framework chosen is Pytorch 7.1.3 and CUDA 12.2.68.

The experimental dataset contains 1697 transmission line foreign object images, with different materials, sizes and shapes, such as bird nests, kites and banners made of fibers such as textiles and paper, balloons or plastic bags made of plastic or rubber, as shown in Fig. 6.

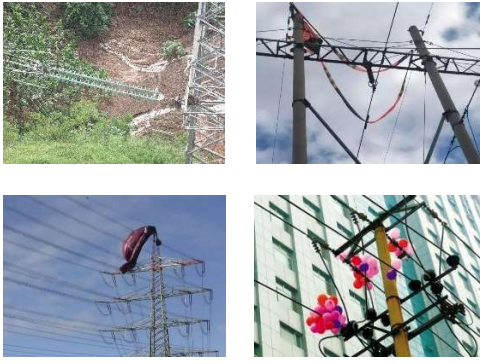

**Fig. 6 Sample pictures of the foreign object in different scenes**

The training dataset was expanded to 2764 images by cropping, scaling, and rotating. The dataset consists of 63.5% bird's nests, 10.8% kites, 14.6% textiles, and 11.1% plastic images, respectively. the dataset without foreign objects and the foreign object dataset has the same image background, they are obtained mainly by cropping the high-resolution foreign object images or masking the foreign objects with background texture. The detail composition of datasets is shown in Table 4.

**Table 4 Number and percentage of pictures of different foreign**

| Type of foreign object | Number of images | percentage share of total number (%) |
|---|---|---|
| bird's nest | 1756 | 63.5 |
| kite | 296 | 10.8 |
| textile | 404 | 14.6 |
| plastic | 308 | 11.1 |
| background | 2800 | -- |

### 5.2 Analysis of recognition accuracy

The evaluation metrics for a model to detect images with foreign objects can be calculated through a confusion matrix, which generally encompasses the following four scenarios:
1) True Positive (TP): When the actual image contains the target, and the detection result also indicates the presence of the target.
2) False Positive (FP): When the actual image is a background or doesn't contain a target object, but the detection result incorrectly indicates the presence of the target.
3) False Negative (FN): When the actual image contains the target, but the detection result incorrectly identifies it as a background or other type of objects.
4) True Negative (TN): When the actual image is a background or contains another type of object, and the detection result also identifies it as not existing.

The evaluation metrics adopted in this paper include the following three:

Recall (True Positive Rate, TPR): It represents the proportion of correctly detected samples containing foreign objects among all positive samples. The formula is:

$$TPR = \frac{TP}{TP+FN}. \quad (26)$$

Precision (Positive Predictive Value, PPV): The proportion of the number of samples actually containing foreign objects to the number of samples containing foreign objects determined by the classifier. The formula is:

$$PPV = \frac{TP}{TP+FP}. \quad (27)$$

Accuracy (ACC): It represents the classifier's ability to judge the entire sample set, i.e., the proportion of correct detections to the total number of samples. The formula is:

$$ACC = \frac{TP+TN}{TP+FP+FN+TN}. \quad (28)$$

The training condition of the improved wavelet scattering deep network is described as follows: the learning rate is set to 0.001, the number of iterations is set to 200, a stochastic gradient descent algorithm is used to update the network weights with a momentum parameter of 0.9. The number of batches for each iteration is 150, and the number of batch samples is 200. In order to validate the performance of our proposed model, SVM, YOLOv5s, and YOLOv8s models are introduced for comparative experiments, where SVM uses a classification function SVC, a linear kernel function is chosen, and the penalty coefficient is 1. The models are used to recognize 3000 randomly selected sample images, which contain 950 foreign object images and 2050 background images, and finally the respective confusion matrix is obtained, and the specific data is shown in Fig. 7.

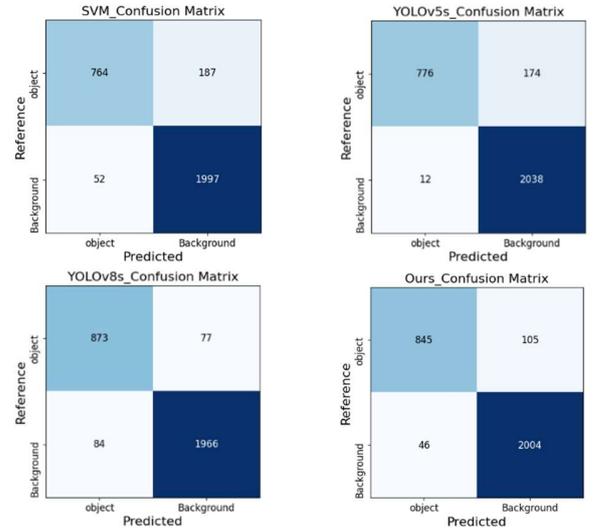

**Fig.7 Confusion matrix obtained from four models of testing**

From Fig.7, we can see that SVM and YOLOv5s have relatively high numbers of FN, which reaches 187 and 174, respectively, indicating that SVM and YOLOv5s are more likely to cause missing detections; YOLOv8s has the highest number of FP, which reaches 84, indicating that YOLOv8s is more likely to cause misdiagnosis; whereas the data of our proposed model are more balanced and exhibits better generalization.

The confusion matrix in Fig. 7, combined with (25), (26) and (27) can be used to find the three evaluation indicators for the four models, and the results are shown in Table 5.

**Table 5 Comparison of recognition data from 4 algorithms**

| Model | TPR/% | PPV/% | ACC/% |
|---|---|---|---|
| SVM | 80.4 | 93.6 | 92.0 |
| Yolov5s | 81.7 | **98.5** | 93.8 |
| Yolov8s | **91.9** | 91.2 | 94.6 |
| Ours | 88.9 | 94.8 | **94.9** |

As can be seen from Table 5, although Yolov8s performs best in recall with 91.9% and Yolov5s performs best in precision with 98.5%, both of them are only four percentage points higher in their respective metrics than our proposed model; and our algorithm has the highest accuracy of 94.9%, which is 1.1% higher than yolov5s and 0.3% higher than yolov8s. Experimental results show that our proposed model exhibits significant advantages, having much lower computational complexity and higher recognition accuracy than other models.

In order to further validate the generalization of our model, detection experiments are conducted for each type of foreign object, where the number of images of each type is randomly divided into training and testing sets in 8:2. Through the experiments, the confusion matrices for various foreign objects recognition can be obtained, as shown in Fig. 8.

**Fig. 8 Confusion matrices for four foreign bodies and backgrounds**

As can be seen from Fig. 8, our proposed model can recognize and classify different types of foreign objects well. Due to the relative complexity of the background of the bird's nest picture, the number of its FP and FN are higher than that of other foreign object types, 26 and 19, respectively, relative to the number of TP and TN, the percentage is less than 10%, the leakage and misdetection rates are low. Combining (25), (26), and (27), we can find the three evaluation indexes of our proposed model for recognizing the four foreign objects, and the resultant data are shown in Fig. 9.

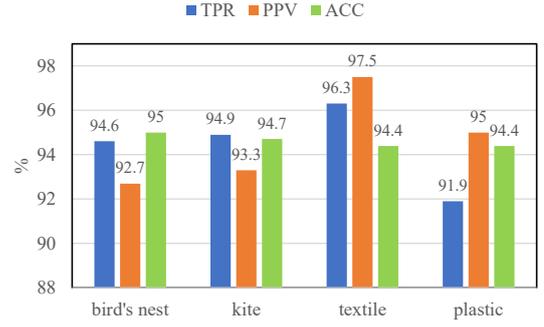

**Fig. 9 Accuracy of different foreign object recognition**

As can be seen from Fig. 9, the recall, precision, and accuracy of our proposed model for recognizing different foreign objects are higher than 90%, among which the recall, precision, and accuracy of recognizing textiles are the highest, 96.3%, 97.5%, and 94.4%, respectively, which indicates that our method is the most effective in recognizing textiles; moreover, the accuracy of recognizing birds' nests is 95%, the accuracy for kites is 94.7%, the accuracy for plastics is 94.4%, and so on. Fig. 9 highlights the high accuracy and generalizability of our proposed model, which has significant application advantages in the field of foreign body detection.

### 5.3 The improvement's impact on performance of Wavelet Scattering Deep Networks

The core differences between the improved and unimproved wavelet scattering networks are the definitions of their high-order module coefficients illustrated and compared in Fig. 10, and these coefficients also serve as the primary features extracted. In the first column of Fig. 10, Figs. 10a, 10d and 10g depict the detected foreign objects, with the first two being bird nests and the last one the balloons. The second column, Figs. 10b, 10e and 10h, show their module coefficients of the improved wavelet scattering network. The third column presents the coefficients of the unimproved one, namely Figs. 10c, 10f and 10i. Among these, the mode coefficients compared in the first and third rows are of the 3rd order, while those in the second row are of the 2nd order.

According to formula (11), the high-order mode coefficients of the unimproved wavelet scattering network are obtained by continuously applying wavelet function transformations to the image, which is equivalent to performing continuous high-frequency feature extraction on the image. This process does not involve the extraction of low-frequency information. In contrast, for the improved nth-order high-order mode coefficients, the first n-1 stages extract low-frequency features, while only the last step extracts high-frequency features. As a result, the extracted results encompass both low-frequency and high-frequency information of the image. Therefore, there are the following differences in the mode coefficient extraction effects between the two:

(1) If there is a significant distinction between the detailed texture features of the foreign object target and the background, the unimproved high-order wavelet mode coefficients will continuously enhance the detailed features of the target. Compared to the improved network's mode coefficients, the former exhibits a more pronounced contrast between the extracted target features and the background image. As shown in Figs. 10b and 10c, due to the substantial differences in the branch details of the bird nest and the tower pole, as well as the leaf details in the background, the contour features of the tower pole and the background are almost indistinguishable in the image of the unimproved wavelet mode coefficients. However, in the improved wavelet mode coefficients, the features of the tower pole and a few background leaves can still be observed.

(2) In scenarios where the foreign object target and the background contain similar detailed features, the unimproved wavelet mode coefficients often struggle to distinguish between the target and the background. In contrast, the improved wavelet mode coefficients can still capture distinct foreign object features because they extract macroscopic characteristic information. As demonstrated in Figs. 10d and 10e or Figs. 10h and 10i, where the branch details of the bird nest almost blend with the weed details in the background, and the balloon foreign object and the background sky are uniformly colored with almost no detailed textures, the improved wavelet mode coefficients can capture the foreign object features, as indicated by the white boxes in the figures. In contrast, the features extracted by the unimproved wavelet coefficients are weak or almost non-existent.

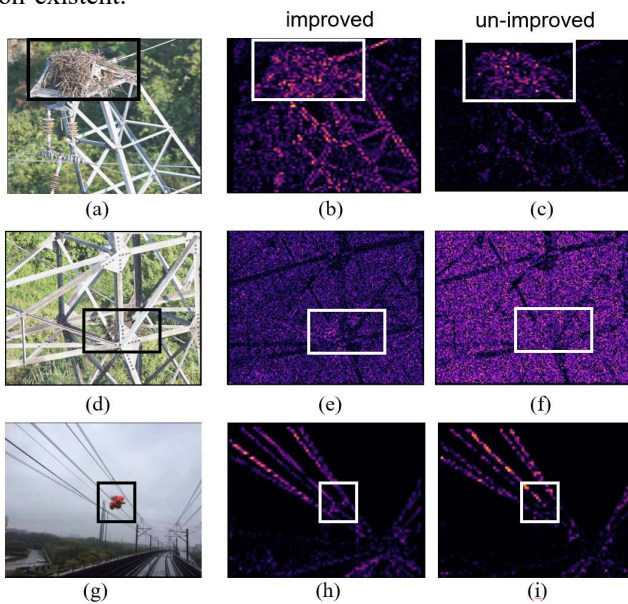

**Fig. 10 Comparison of feature extraction results. (a)(b)(c) Nest image and the 3$^{rd}$ order modulus coefficients of the improved or un-improved scattering network. (d)(e)(f) Nest images and its 2$^{nd}$ modulus coefficients, in which object has similar textile to the background. (g)(h)(i) Ballen images and 3$^{rd}$ order modulus coefficients of two type networks.**

Above analyses on objects extracted features indicates that both improved and original scattering deep networks are more beneficials to detect certain type of foreign objects. To further compare their total performance and robustness in detail, the entire foreign object detection experiments are also conducted using the unimproved wavelet scattering deep networks, as the results are discussed as follows.

Table 6 presents the recognition accuracy and inference speed of the unimproved wavelet scattering deep network for various foreign objects. The values within parentheses denote the performance variances relative to the improved model. For objects such as bird nests and textiles, most performance indicators of the unimproved network, including TPR, PPV and ACC, are lower than those of the improved network. Notably, bird nests are often the primary detected objects that lead to larger transmission line faults. The accuracy of the unimproved network for plastic objects remains comparable to that of the improved network. However, the unimproved network exhibits superior accuracy for kite foreign objects. The inference time of the unimproved network is 6.7 ms, which is 0.2 ms longer than that of the improved network.

**Table 6 Performance of un-improved model and its differences from the improved one (in parenthesis)**

| Type | TPR/% | PPV/% | ACC/% | Speed/ms |
|---|---|---|---|---|
| Nest | 88.3 (-6.3) | 92.5 (-0.2) | 92.0 (-3.0) | |
| Kite | 98.3 (+3.39) | 90.6 (-2.7) | 96.0 (+1.3) | 6.7(0.2) |
| Textile | 95.1 (-1.2) | 98.7 (+1.2) | 90.0 (-4.4) | |
| Plastic | 95.2 (+3.2) | 95.2 (+0.2) | 94.5 (+0.1) | |

Fig. 11a presents the confusion matrix of the unimproved wavelet scattering deep network for different foreign objects. Fig. 11b shows the difference between the confusion matrices of the improved and unimproved models, and its diagonal elements are highlighted with blue boxes. The sum of the diagonal elements in Fig. 11b is greater than 0, indicating that for the entire dataset, the precision (TPR and PPV) of the improved algorithm is higher than that of the unimproved one. In the case of bird's nest foreign object incidents, the accuracy of the unimproved wavelet scattering deep network is significantly lower than that of the improved one. Considering both the overall recognition accuracy and inference speed, the improved wavelet scattering deep network exhibits better comprehensive performance.

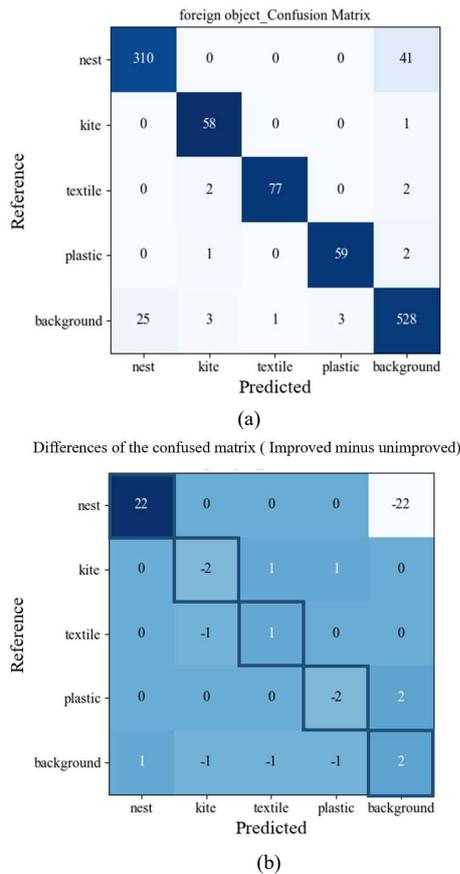

**Fig. 11 Confusion matrices for four foreign bodies and backgrounds. (a) Confusion matrix of un-improved models, (b) The improved models' matrix minus the un-improved one, namely their differences.**

### 5.4 Analysis of Inference Efficiency for Foreign Object

In order to verify the applicability and real-time performance of our algorithm, the model is deployed to five different edge devices for experiments, including Raspberry Pi 4B produced by Raspberry Pi Foundation, Orange Pi 3B, Orange Pi Zero3 and Orange Pi 5B produced by Orange Pi, Inc. Jetson Nano produced by NVIDIA. The specific object is shown in Fig. 12.

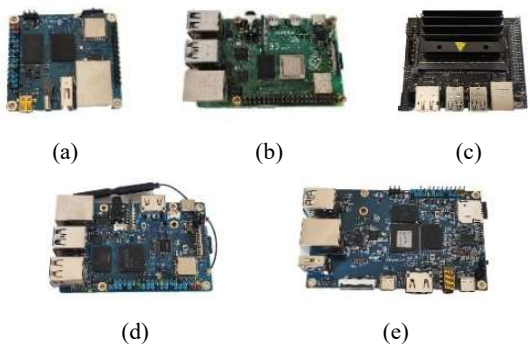

**Fig. 12 Physical drawings of the five edge devices, (a) Orange Pi Zero3, (b)Raspberry Pi 4B, (c) Jetson Nano with 2GB graphic memories, (d) Orange Pi 3B and (e) Orange Pi 5B.**

The performance parameters of the five devices are shown in Table 7, where the Orange Pi 5B has the highest peak arithmetic power of 3000 GFLOPS with ARM Mali-G610 MP4 CPU; the Orange Pi Zero3 has the smallest peak arithmetic power of 333 GFLOPS with ARM Mali G31 MP2 CPU; the Jetson Nano contains a 128-core Maxwell™ GPU, and the ARM processors of the other devices also contain high-performance image processing units. In addition, Table 7 shows the detection speed of each edge device for 1280×720 resolution images, as well as the efficiency of their hardware optimizations. The hardware optimization efficiency is defined as:

$$efficiency = \frac{FPS}{FLOPs}, \qquad (29)$$

where *FPS* is the detection speed and *FLOPs* is the peak arithmetic power of the device. Hardware optimization efficiency reacts to the matching degree and optimization efficiency between hardware and algorithm. The higher the matching degree, the higher the actual inference performance under the same arithmetic index.

**Table 7 Five edge device parameters and their inference speed**

| edge device name | CPU/frequency/core | FLOPS | FPS | efficiency |
|---|---|---|---|---|
| Orange Pi Zero3 | full-fledged H618 Cortex-A53/1.5GHz/4 | 333G | 37.0 | 0.111 |
| Orange Pi 3B | RMC RK3566/1.8GHz/4 | 400G | 36.4 | 0.091 |
| Nvidia Jetson Nano | ARM Cortex-A57/1.43GHz/4 | 472G | 66.7 | 0.141 |
| Raspberry Pi 4B | ARM Cortex-A72/1.5GHz/4 | 576G | 39.5 | 0.069 |
| Orange Pi 5B | RMC RK3588S/2.4GHz/8 | 3000G | 149.3 | 0.050 |

As can be seen from Table 7, our proposed model detects 1280×720 HD images at a speed greater than 30 FPS on all five edge devices, with the fastest detection speed reaching 149.3 FPS on Orange Pi 5B, and 36.4 FPS on Orange Pi 3B, which has the relatively lowest performance and power consumption. In addition, the detection speed of the model shows a positive correlation with the CPU and GPU performance of the edge device, and the relationship between its peak arithmetic power and detection speed is shown in Fig. 13.

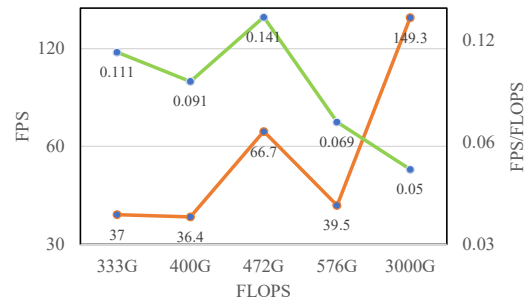

**Fig.13 The relationship between peak computing power and detection speed**

As seen in Fig. 13, the Jetson Nano has the highest degree of hardware optimization, which can reach 0.141 FPS/FLOPS, compared to other devices with similar arithmetic power, such as the Raspberry Pi 4B, Orange Pi 3B, and Orange Pi Zero3, its inference speed is nearly twice as fast as that of these three, which indicates that the Jetson Nano uses a better performing GPU. The Orange Pi 5B has the fastest detection speed and highest performance, but it has the least efficient hardware optimization at 0.05 FPS/FLOPS. The Orange Pi Zero3 has the lowest inference performance, but it has a level of hardware optimization for the algorithm that is second only to the Jetson Nano.

### 5.5 Effect of Image Input Resolution on Detection Speed

The effect of image resolution on the detection speed of the algorithm is shown in Table 8, and the experimental equipment is Orange Pi 5B. The table gives the total computational complexity of the model, theoretical inference time, and actual inference time for images of different resolutions such as 540P, 720P and 1080P. The contents of the table are presented as a matrix drawing, where the inference time is converted into FPS, as shown in Fig. 12.

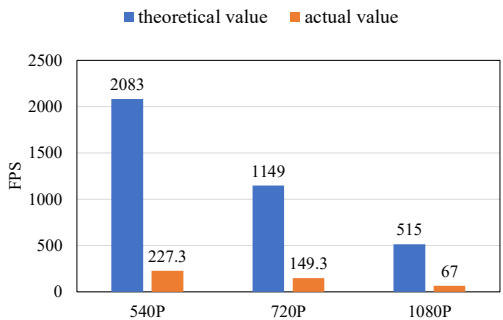

**Fig. 14 Rectangular plots of theoretical and actual values of inference speed at different resolutions**

As can be seen from Fig. 14, the inference speed of our proposed model reaches 232.6 FPS for 540P images, which is dozens of times faster than that of a large model with the same the input image size. Furthermore, the speed also reaches 67 FPS at 1080P resolution, which proves that the method can be used to perform real-time on-line detection of super-high-resolution images.

**Table 8 Comparison of computational complexity and corresponding inference time for different resolutions**

| image size | FLOPs | theoretical inference time (ms) | actual inference time (ms) |
| --- | --- | --- | --- |
| 960×540 | 45.4M | 0.48 | 4.4 |
| 1280×720 | 81.1M | 0.87 | 6.7 |
| 1920×1080 | 181.7M | 1.94 | 14.9 |

In addition, from Table 8, we found that the theoretical or practical inference time is directly proportional to the resolution of the image, which proves the accuracy of the model's computational complexity inference equation (23). And there is a proportional relationship of about 10 times between the actual inference time and the theoretical inference time due to the two reasons: 1) the implementation of the algorithm is based on python language, which is less efficient than the static programming languages such as C and Rust; 2) the Orange Pi 5B hardware itself is not optimized enough for the algorithm to be the most efficient, as shown in Table 8, which is the lowest among all the edge devices. Therefore, our proposed algorithm has a large optimization space in both software implementation and hardware optimization efficiency, and the inference time can be further improved in the future.

### 5.6 Performance Comparison of Other Algorithms

We compared our model's performance with methods proposed in existing references, as shown in Table 9.

**Table 9 Performance comparison of different algorithms**

| reference | image size | GPU detection speed/FPS | Edge device detection speed /FPS | ACC/% |
| --- | --- | --- | --- | --- |
| (Hu et al., 2024) | multi-scale | -- | 79 | 92.3 |
| (Lu et al., 2020) | 300×300 | 101.2 | -- | 89 |
| (Liu et al., 2024) | 640×640 | 277.8 | 142.9 | 75.6 |
| (Niu et al., 2023) | 416×614 | 52 | -- | 90.1 |
| (Lu et al., 2023) | 640×640 | 111 | 63 | 94.3 |
| (Han et al., 2024) | 640×640 | -- | 58 | 89.2 |
| (Han et al., 2023) | 640×640 | **285.7** | 23.5 | 93.7 |
| Ours | **1280×720** | 158.7 | **149.3** | **94.7** |

As can be seen from Table 9, although the input image size of our algorithm is 720P, which is a higher resolution than all other models, the detection speed on the edge device is the fastest among so models, reaching 149.3 FPS, which is double that of the LEE-YOLOv7 model (Hu et al., 2024), and triple that of the YOLO-GSS model (Han et al., 2024); Mobilenet+SSD (Lu et al., 2020) and Comprehensive YOLOv5 (Niu et al., 2023) can be deployed on edge devices, but their detection speeds are significantly lower than that of our algorithm; the reparameterized YOLOv5 (Liu et al., 2024) and TD-YOLO (Han et al., 2023) detect very fast on GPU, reaching 277.8 FPS and 285.7 FPS, respectively, which are about twice as fast as our algorithm, but their speeds decreases drastically when deployed to the edge devices, whereas our method suffers from a very small impact; Although the YOLO-inspection accuracy (Lu et al., 2023) is the highest among the references, at 94.3%, it is slightly lower than our algorithm's 94.7%, and its detection speed is lower than that of our algorithm both on the GPU and on edge devices. Therefore, our proposed algorithm ensures the recognition accuracy with high real-time performance and enables local real-time detection of HD video streams.

## 6 Conclusion

The proportion of renewable energy generation in the total power generation is increasingly high, which poses higher real-time requirements for the operation and maintenance of transmission lines. This paper proposes a real-time foreign object recognition model based on an improved wavelet scattering deep network. The model combines the advantages of wavelet scattering network and neural network with the following features: firstly, a wavelet scattering network is utilized instead of the convolutional and pooling layers in the CNN, which greatly reduces the computational effort of the model; secondly, the structure of the wavelet scattering network is optimized, and the biorthogonal wavelets basis including bior1.1, bior2.2, and bior1.3 wavelet are utilized, which can effectively extract the features of foreign objects; finally, the extracted modulus and scattering coefficients feature combinations are recognized using a 3-layer mini-deep network, which significantly reduces the amount of information input compared to the original high-resolution image and significantly improves the detection speed and accuracy. The experimental results show that the model proposed in this paper has a detection accuracy of more than 90% for 1280×720 images. In comparative experiments with the three models of SVM, YOLOv5s, and YOLOv8s, our model yields the highest accuracy rate of 94.9%, 1.1% higher than YOLOv5s and 0.3% higher than YOLOv8s; In addition, the model is deployed on five edge devices with detection speeds of more than 30 FPS. And 149.3 FPS can be reached on the Orange Pi 5B, enabling local real-time detection of HD video streams.

The method proposed in this paper can efficiently accomplish the task of real-time detection of foreign objects on edge equipment, which is valuable in practical engineering application cases.

**Acknowledgments**

This work is supported by the National Natural Science Foundation of China (No. 51807030) and Guiding Project of Fujian Provincial Department of Science and Technology (No. 2017H0013)

**Author contributions**

Nan Xie designed the research, write the artificial intel-ligence programs and organized the manuscript. Zhi-chao He conducted the experiments and drafted the manuscript. Xiang-yu Shen and Zhang Yong help to conduct the experi-ment and revised the paper. Nan Xie and Zhi-chao He revised and finalized the paper.